\documentstyle[prl,aps,epsf,floats]{revtex}

\newcommand{\be}{\begin{equation}}              
\newcommand{\ee}{\end{equation}}                
\newcommand{\bea}{\begin{eqnarray}}             
\newcommand{\eea}{\end{eqnarray}}
\newcommand{\nn}{\nonumber}

\newcommand{\rr}{{\bf r}}
\newcommand{\qq}{{\bf q}}

\newcommand{\GG}{{\bf G}}

\begin{document}

\flushbottom

%---------------------------------------------------------------- 

\draft
\wideabs{              % makes abstract and title in one column
\vspace{-.4cm}
\title{Disclination Unbinding Transition in Quantum Hall Liquid Crystals
\vspace{-.3cm}
}
\author{C. Wexler$^1$ and Alan T.\ Dorsey$^2$}
\address{$^1$Department of Physics and Astronomy, 
         University of Missouri--Columbia, 
         Columbia, Missouri 65211}
\address{$^2$Department of Physics, University of Florida,
         P.O. Box 118440, Gainesville, Florida 32611-8440}
\date{18 October 2000}
\maketitle

\vspace{-.2cm} 

\begin{abstract}
We derive the the long-wavelength elastic theory for the quantum Hall 
smectic state starting from the Hartree-Fock approximation.  Dislocations
in this state lead to an effective nematic model for $T>0$, 
which undergoes a disclination unbinding transition from a phase with algebraic
orientational order into an isotropic phase.  We obtain transition 
temperatures  which are in qualitative agreement with recent experiments 
which have observed large anisotropies of the longitudinal 
resistivities in half-filled Landau levels, 
lending credence to the liquid crystal interpretation of experiments.
\end{abstract}

\vspace{-.3cm} 

\vspace{-.55cm}
}                      % end of wideabs

%\begin{multicols}{2}

%-----------------------------------------------------------------
Recent experiments \cite{lilly99a,du99,shayegan99} in high mobility
two-dimensional electron systems (2DES) have revealed remarkable new
phenomena in the transitional regions between the different plateau of
the Hall conductance.  In particular, striking anisotropies and
non-linearities in the magnetotransport were observed for Landau level
(LL) filling factors near $\nu = n + 1/2$, for $n \ge 4$,
corresponding to partially filled  LL indices $L \ge 2$.  This
anisotropy tends to align with the crystalline axes of the sample, but
can be reoriented by the application of in-plane magnetic fields
\cite{pan99,lilly99b}, and resistance ratios as high as 
$R_{xx}/R_{yy} \sim 3500$ have been observed \cite{simon99}.  
This anisotropic behavior has been attributed to the formation of a
striped phase.  A unidirectional charge density wave (UCDW) had been
predicted several years ago \cite{foglermoessner} for nearly half
filled high LLs; exact diagonalizations for systems of up to
12 electrons \cite{rezayi} corroborate this picture for $L \ge 2$, and 
many experimental results can be qualitatively understood under the
assumption of a UCDW.  The presence of stripes has already been
directly observed in a large class of low-dimensional, strongly
correlated electronic systems \cite{tranquada95}, and the present 
experimental evidence in quantum Hall devices is compelling, even if
still somewhat circumstantial \cite{eisenstein00a}.

Due to the similarities of the UCDW state with a classical smectic
liquid crystal, these states have been  dubbed 
{\em quantum Hall smectics} by Fradkin and Kivelson 
\cite{fradkin99,fradkin00}.  In two dimensions thermal
fluctuations destroy the positional order \cite{zero_temperature}, 
but the system should still exhibit anisotropic transport as long as
there is some remnant of orientational order  (algebraic order in the 
{\em quantum Hall nematic}) \cite{toner81}.  As the temperature is
increased, the algebraic orientational order will disappear in a
Kosterlitz-Thouless (KT) disclination-unbinding transition
\cite{kosterlitz73}.   

To study this process we have mapped the interacting electron system
(in the Hartree-Fock approximation) onto a classical smectic (the
UCDW).  We then consider the role of thermal fluctuations (phonons and
dislocations) in reducing the order from smectic to nematic at larger
distances.   Without the use of any fitting parameters, and using only
experimentally accessible values for the electron density and the
width of the 2DES, we are able to estimate values for the disclination
unbinding transition temperature, which are in qualitative  agreement
with the transport measurements. 

%-----------------------------------------------------------------
{\em (I) Hartree-Fock approximation for the charge-density-wave
  state.---}In order to study the energetics of a charge density wave 
(CDW) in the 2DES we closely follow the strategy developed in
Refs. \cite{macdonald84,jungwirth99,stanescu00}, and use the
Hartree-Fock (HF) approximation, which corresponds to the assumption
that the electronic state can be described as a Slater determinant of
single-electron states.  In the Landau gauge, 
${\bf A}({\bf r}) = (0,B x,0)$, and the eigenstates of the
non-interacting problem are 
\be
\label{eq:non-interacting-states}
\psi_{\alpha \sigma n x_0} ({\bf r}) =
         \frac{\zeta_\alpha (z) \, e^{i x_0 y/l_b^2} 
        \, H_n\bigl(\frac{x-x_0}{l_b}\bigr) \, e^{-(x-x_0)^2/2 l_b^2}}
        {\pi^{1/4} (2^n n! L_y)^{1/2}} \,,
\ee
where $\alpha$, $\sigma$, $n$ and $x_0$ indicate the electric sub-band
index (due to the confinement in the $z$ direction), spin index, LL
index, and guiding center respectively; $l_b = (\hbar/e B)^{1/2}$ is
the magnetic length, $L_y$ is the length of the system in the $y$ 
direction, and $H_n$ are Hermite polynomials. 

Since the electric sub-band splitting is very large (about 9.8 meV in
the sample of Ref. \onlinecite{lilly99a}), in what follows we consider
only states with $\alpha = 0$.  The Coulomb interaction between the
basis states above can be replaced by the effective
interaction \cite{macdonald84,jungwirth99,stanescu00}
\be
\label{eq:Vq-eff}
V^{n_1, n_2}_{x_1, x_2}(q_x,q_y) =  \frac{4 \pi e^2}{\kappa} \int  dq_z \; 
        \frac{|M^{n_1, n_2}_{x_1, x_2}(\qq)|^2 }
                {q^2} \,,
\ee
where $\kappa$ is the dielectric constant of the semiconductor ($\sim \!\!
13$ in GaAs/AlGaAs), with the matrix element
\be
\label{eq:matrix-element}
M^{n_1, n_2}_{x_1, x_2}(\qq) = \int d^3x \; e^{i \qq \cdot \rr} \,
        \psi_{0 \sigma n_1 x_1}^* ({\bf r})
        \psi_{0 \sigma n_2 x_2} ({\bf r})
\ee
which may be expressed in terms of associated Laguerre polynomials
\cite{macdonald84,jungwirth99,stanescu00}. 
Since the anisotropic states occur for moderately weak magnetic
fields, the effect of a CDW on the valence LL is to polarize the fully
occupied LLs below.  This polarization may be accounted for with an effective
dielectric constant $\epsilon({\bf q})$, which can be 
calculated in the random phase approximation (RPA)
\cite{jungwirth99,stanescu00,manolescu97}. This effective interaction
greatly simplifies the calculation,  as we only need to consider
states within the valence LL for the determination of CDW energies.  

\begin{figure}[t]
\vspace{-.2cm}
 \begin{center}
   \leavevmode
  \epsfbox{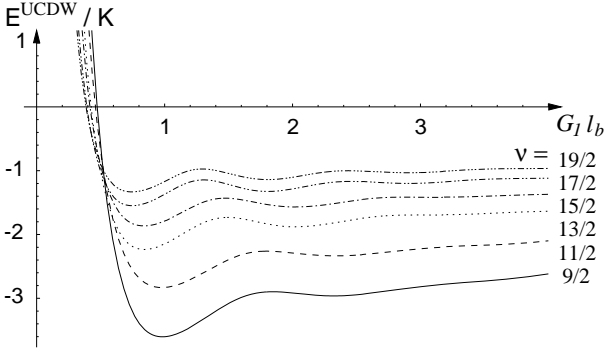} \vspace{.0cm}
   \caption{\label{fig:E-UCDW} 
       Dependence of the average energy per electron state 
       $E^{\rm ucdw}$ for various filling factors of 
       Ref.\ \protect\onlinecite{lilly99a}
        ($n_e = 2.67 \times 10^{11}$ cm$^{-2}$, 
        $z_{\rm rms} = 58.3 {\; \rm \AA}$).
      }
 \end{center}
\vspace{-.75cm}
\end{figure}

In the absence of LL mixing, the state of the system is uniquely
specified by the particle density function
\cite{macdonald84,macdonald88}.  The energy per electron in a CDW
state at a fractional filling $\nu^*$ is given by \cite{wexler00a}
\be
\label{eq:cdw-energy}
E = \frac{1}{2 \nu^*} \sum_j 
  U(\GG_j) \,  |\Delta (\GG_j)|^2  ,
\ee
where $\Delta (\GG_j)$ is the Fourier coefficient of the occupation number
at the reciprocal lattice vector $\GG_j$ and the kernel 
$U(\qq)=H(\qq)+X(\qq)$ with the direct and exchange contributions
\bea
\label{eq:direct}
H(\qq) & = & \frac{1}{2 \pi l_b^2 \, \epsilon(q)} 
V^{n,n}_{x_1,x_1+l_b^2 q_y}(q) \,, \\
\label{eq:exchange}
X(\qq) &=& - \int \frac{d^2p}{(2 \pi)^2 \, \epsilon(p) } \,
    e^{i(p_x q_y - p_y q_x)l_b^2} \, 
V^{n,n}_{x_1,x_1+l_b^2 p_y}(p) .
\eea
%
%where $V^{n,n}(q) \equiv V^{n,n}_{x_1,x_1+l_b^2 q_y}$. 
In the UCDW state, we have $\GG_j = {\bf e}_x \, G_1 \, j \,$ with 
$j$ an integer, and
\be
\label{eq:UCDW-Delta}
\Delta(\GG_j) = \frac{ \sin (\nu^* \pi \, j)}{ \pi \, j} \,,
\ee
where  $G_1 = 2 \pi/a$, with $a$ the period of the UCDW.  Inserting
this into Eq.\ (\ref{eq:cdw-energy}) we find $E^{\rm ucdw}(G_1)$, the
average energy per electron in the UCDW (see Fig.\ \ref{fig:E-UCDW}).
The optimal UCDW corresponds to the minimum $E^{\rm ucdw}$, and is
observed at $a \simeq 2.84 \, l_b \sqrt{2 L + 1}$ (in general 
agreement with Ref.\ \onlinecite{foglermoessner}, even though we are
far from $L\!\rightarrow \! \infty$), where each electron
gains one to a few degrees, see Table \ref{tabl:HF-elastic}. Since the
anisotropic-isotropic transition is  observed at temperatures much
smaller than this, it is clear that the observed transition is not
related to the formation of the stripes but, as we shall see, to the
unbinding of topological defects in the stripes. 

\begin{figure}[t]
 \begin{center}
   \leavevmode
   \epsfbox{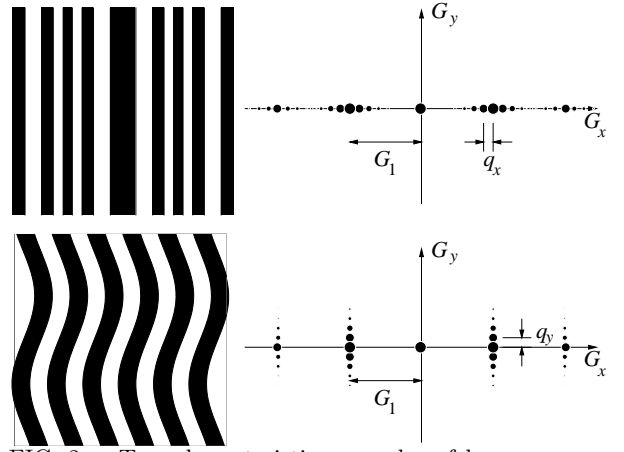} \vspace{.0cm}
   \caption{\label{fig:MODES}
     Two characteristic examples of low energy perturbations of the
     UCDW.  Top: the {\em longitudinal modulation}.  Bottom: the
     {\em transverse modulation}.  On each panel, the
     right-hand side shows the Bragg peaks of $\Delta(\GG)$ in
     reciprocal space. $G_1$ is the wavevector of the UCDW, and $q_x$, 
     $q_y$ are the wavevectors of the modulation.  See Eqs.\
     (\ref{eq:distortion-u})--(\ref{eq:bragg2}).
     }
 \end{center}
\vspace{-.7cm}
\end{figure}

%-----------------------------------------------------------------
{\em (II) Low energy excitations of the UCDW.---}Here we consider low
energy states which correspond to long wavelength fluctuations of the
UCDW.  We take care to construct modulations of the stripes which do
not accumulate charge over large distances since this would
significantly increase the Coulomb energy of the system.
These modulations add  extra ``Bragg peaks'' to the density function
$\Delta(\GG)$ (see Fig.\ \ref{fig:MODES}), and of the many modulations
one can devise, very few avoid adding significant peaks far from where
$U(\GG)$ is near its minimum \cite{wexler00b}.  These can be described
by a distortion in the position of the UCDW stripe edges of the form
\be
\label{eq:distortion-u}
u(x,y) = \alpha \, \cos(q_x \, x) \, \cos(q_y \, y) ,
\ee
where $\alpha$, $q_x$, $q_y$ are the amplitude and wavevector
components of the modulation respectively.  
%This perturbation in 
%$u(x,y)$ corresponds to a series of ``Bragg peaks'' of $\Delta(\GG)$ in
%reciprocal space.  
Longitudinal ($q_y=0$) and transverse ($q_x=0$)
modulations are illustrated in Fig.\  \ref{fig:MODES}.  To
determine the energy of this excited state to ${\cal O}[\alpha^2]$, we
need to retain the following peaks:
\bea
\label{eq:bragg1}
&& \Delta[{\bf e}_x  G_1 j ] =
\frac{\sin (\nu^* \, \pi \, j)}{\pi \, j} 
\left( 1 - \frac{G_1^2 j^2 \alpha^2}{8}\right), \\
\label{eq:bragg2}
&& \Delta[{\bf e}_x (G_1 j \pm  q_x) \pm {\bf e}_y q_y] =
-  \frac{\sin (\nu^* \pi j)}{\pi j} \frac{G_1 j \, \alpha}{4} ,
\eea
where $j$ is an integer.  The energy per electron, relative to the
optimal UCDW is then given by
\bea
\Delta E &=& {G_1^2 \alpha^2 \over 16\pi^2 \nu^*}
\sum_{j=-\infty}^{\infty} 
	\sin^2 (\nu^* \pi j)\left[
	U \Bigl( \sqrt{(G_1 j + q_x)^2 + q_y^2} \, \Bigr)
\right. \!\!\!\!
\nn \\
&& \left. 
	+ \, U \Bigl(  \sqrt{(G_1 j - q_x)^2 + q_y^2} \, 
	\Bigr) -2 U(G_1 j) \right] .
\eea
Keeping terms up to ${\cal O}[q_x^4,q_y^4,q_x^2q_y^2]$, the energy
{\em per unit area} is  
\be
\label{eq:energy_per_area}
\Delta {\cal E} = \frac{\alpha^2}{8} \bigl[ {B} q_x^2 + {K} q_y^4 
	+ {K'} q_x^2 q_y^2 + {K''} q_x^4 \bigr] ,
\ee
with the elastic coefficients given by 
\bea
\label{eq:B}
{ B} &=& \frac{\nu^*}{2 \pi l_b^2} \frac{G_1^2 \partial^2 E^{\rm
ucdw}}{\partial G_1^2} , 
\\
\label{eq:K}
{ K} &=& \frac{1}{16 \pi^3 l_b^2} \!\sum_{j=-\infty}^{\infty} \!\!
	\frac{\sin^2(\pi \nu^* j)}{j^2} \Biggl[
	U''(G_1 j) - \frac{ U'(G_1 j)}{G_1 j}\Biggr] \!,\!\!\!\!
\\
\label{eq:K1}
{ K'} &=& 
	{G_1\over 4 \pi^3 l_b^2}\sum_{j=-\infty}^{\infty} 
 \frac{\sin^2(\pi \nu^*j)}{j} 
	\Biggl[ \frac{U'''(G_1 j)}{2} 
\nn \\ && \hspace{2cm}
   	- \frac{ U''(G_1 j)}{G_1 j} + \frac{ U'(G_1 j)}{G_1^2 j^2}
	\Biggr], 
\\
\label{eq:K2}
{ K''} \!&=& \frac{G_1^2}{48 \pi^3 l_b^2} \sum_{j=-\infty}^{\infty} 
	{\sin^2(\pi \nu^* j)} \,	U''''(G_1 j) .
\eea

It is easy to see from energetics above [Eq.\
(\ref{eq:energy_per_area})] that the
low-energy perturbations of a UCDW correspond one-to-one to those of a
smectic liquid crystal \cite{toner81}: 
\bea
\label{eq:LCmacro}
E_{\rm sm} &=& \frac{1}{2} \int d^2 r \, \Bigl\{ \left[
        { B} \, (\partial_x u)^2 
	+ { K} \, (\partial_y^2 u)^2 \right] 
\nn \\ && \hspace{1.8cm}
	+ \left[ { K'} \, (\partial_x \partial_y u)^2
	+ { K''} \, (\partial_x^2 u)^2  \right]  \Bigr\} .
\eea
Results for the elastic moduli ${ B}$ and ${ K}$ are presented 
in Table \ref{tabl:HF-elastic} for parameters relevant to the sample
used in Ref\ \onlinecite{lilly99a}.  The terms between the second sets 
of brackets in Eq.\ (\ref{eq:LCmacro}) are not expected to be 
relevant since they only become large for momenta near the 
edge of the Brillouin zone (where the
validity of the elastic theory is doubtful).  

We now use the energy functional $E_{\rm sm}$ (without the terms
involving ${ K'}$ and ${ K''}$) for all further analysis
of the quantum Hall liquid crystal.

\begin{table}[t]
\vspace{-.3cm}
\begin{center}
     \caption{   \label{tabl:HF-elastic} 
        UCDW:  optimal wavevector $G_1$, period $a$, energy
        gain per electron $E^{\rm ucdw}$ and elastic
        constants $B$ and $K$.  
        The calculations were performed for the
        realization of Ref.\ \protect\onlinecite{lilly99a}.
      }
   \vspace{.0cm}
\begin{tabular}{|r|r|r|r|r|r|r|r|}
\hline
$\nu$ & $\!B$(T) & $l_b\!\!$ (\AA) & $G_1 \, l_b$ & $a\!\!$ (\AA) 
& $E^{\rm ucdw}\!\!$ (K) 
& $B { ( \mu {\rm K}/{\rm \AA}^2)}$ 
        & $\!K$(mK)$\!$  \\
\hline
9/2  & 2.46 & 164 & 0.983 & 1048 & -3.603 
        & 25.5 & 189 \\ %\hline
11/2 & 2.02 & 181 & 0.978 & 1163 & -2.830 
        &  15.7 & 144 \\ %\hline
13/2 & 1.70 & 197 & 0.842 & 1470 & -2.234 
        &  13.0 & 192 \\ %\hline
15/2 & 1.48 & 211 & 0.839 & 1580 & -1.864 
        & 9.07 & 158 \\ %\hline
17/2 & 1.30 & 225 & 0.746 & 1895 & -1.549 
        & 7.58 & 196 \\ %\hline
19/2 & 1.16 & 239 & 0.744 & 2018 & -1.332 
        & 5.66 & 167 \\ 
\hline
\end{tabular}
\end{center}
\vspace{-.7cm}
\end{table}

%-----------------------------------------------------------------
{\em (III) Effects of thermal fluctuations: 
from smectics to nematics.---}The energy functional for a smectic 
[Eq.\ (\ref{eq:LCmacro})] has been extensively
studied.  We follow closely the formulation of Toner and Nelson
\cite{toner81}.  Since the dimensionality of the system ($d=2$) is one
below the lower critical dimension for layered materials, phonon
fluctuations readily destroy positional order for $T>0$
(the Landau-Peierls argument), while preserving 
order in the layer orientation.
However, this argument omits dislocations, which 
have finite energy; their energy can be estimated as \cite{wexler00b} 
\be
E_D = \frac{  B a^2 }{ 4 \pi} 
        \bigl[ \sqrt{2 q_c \lambda+1} - 1 \bigl]\,,
\ee
where $\lambda^2 = { K/B}$ and $q_c \!\sim\! \pi/a$ is a large
momentum cut-off.  Therefore, for $T>0$ we expect 
a density of dislocations given by 
$ n_D \approx a^{-2} \, e^{-E_D/k_B T}.$
At distances larger than $\xi_D = n_D^{-1/2}$, and as long as 
$E_D \! \not \gg \! k_B T$, dislocations can be
treated in  a Debye-H\"uckel approximation.  Then, to lowest order in
$q_x^2$ and $q_y^2$, the correlation function for the 
layer normal angle $\theta = - \partial_y u$ can be written
as \cite{toner81} 
\be
\langle \tilde{\theta}(\qq) \tilde{\theta}(-\qq) \rangle 
 = \frac{ k_B T }{ 2 E_D \, q_x^2 +  K q_y^2 } \,,
\ee
which is precisely the correlation function of a two-dimensional
nematic, with a free energy 
\be
\label{eq:F-nematic}
F_{\rm nm} = \frac{1}{2}  \int \! d^2r  \Bigl[
        K_1 ( \nabla \!\cdot \!{\bf n} )^2 + 
        K_3\bigl[ {\bf n} \!\times\! 
(  \nabla \!\times\! {\bf n} ) \bigr]^2
\Bigr] ,
\ee
where ${\bf n} = (\cos \theta, \sin \theta)$ is the director field,
and the two Frank constants are given by
\be
\label{eq:K1-K3}
K_1 =  K\;\;\; {\rm and} \;\;\; K_3 = 2 E_D \,.
\ee
Orientational correlations in the director ${\bf n}(\rr)$ should decay 
algebraically at distances much larger than $\xi_D$.  Table
\ref{tabl:sm2nem2kt} summarizes the values of $K_1$ and $K_3$. The
values of these elastic constants are determined at distances
comparable to $\xi_D$ ($\sim 10 \, a$ at $T \sim 100$ mK).

%-----------------------------------------------------------------
{\em (IV) The nematic to isotropic transition.---}At
sufficiently long wavelengths Nelson and Pelcovits \cite{nelson77}, 
using a momentum-shell renormalization approach, 
have shown that deviations from the one-Frank constant approximations
$K_1=K_3$ are irrelevant, and the system is equivalent to a 
two-dimensional XY model:
\be
\label{eq:F-xy}
F_{\rm xy} = \frac{1}{2} \, K(T) \int d^2r \, (\nabla \theta)^2 , 
\ee
with $K \rightarrow [ K_1(\xi_D) + K_3(\xi_D) ]/2$ at very large
distances.  For our values of $K_1$ and $K_3$, at the characteristic
temperatures of the experiments, convergence is achieved at distances
around  20--100 $\xi_D$. 
We then expect unbinding of disclination
pairs at the KT temperature \cite{kosterlitz73}: 
\be
k_B T_{KT} = \frac{\pi}{8} \, K(T_{KT}) \,,
\ee
where the $\pi/8$ comes instead of the more common $\pi/2$ for
vortices since each disclination winds up the angle by $\pi$ rather
than $2\pi$.  In general, $K(T_{KT})$ corresponds to the
large-distance elastic constant (reduced by disclination pairs) to the
bare  elastic constant at small distances $K(0)$ by means of the 
KT RG formulas \cite{kosterlitz73}:
\be
\label{eq:RG-KT}
%\left\{ 
%\begin{array}{l} 
%\displaystyle
    \frac{d k^{-1}}{d l}  =  \pi^3 y^2(l) \,, \;\;\;\;\;
%\displaystyle
    \frac{d y}{d l}  = [8 - \pi k(l)]\, \frac{y(l)}{4} \,,
%\end{array} \right.
\ee
where $k = K/k_BT$ and we have introduced the fugacity 
$y \!\sim\! \exp[-\pi^2 \, K(0)/k_B T]$.  In practice, these RG equations
can be approximated by $ k_B T_{KT} \simeq (\pi/8) \, K (0) / 
(1 \!+\! 2 \pi \exp[-\pi^2 K(0)/8 k_B T_{KT}]) 
\simeq 0.86 \, (\pi/8) \, K (0)$. This reduction is in general
agreement (although somewhat less important) to 
results for Monte-Carlo simulations \cite{fradkin00}.  

Table \ref{tabl:sm2nem2kt} presents the resulting
estimates for the disclination unbinding transition temperatures for
half-filled LLs.  Although these can only be considered estimates due
to the approximations used, they are in qualitative agreement with
the temperatures at which the anisotropies are seen to vanish.  We also
see the characteristic spin oscillation of the transition parameters
\cite{lilly99a,eisenstein00b}.   
The reason for this spin oscillation is simple:  in the
energetics of Eqs.\ (\ref{eq:cdw-energy}-\ref{eq:exchange}), there is
an energy scale $e^2/l_b$ that decreases with increasing filling
factor $\nu$; simultaneously the matrix elements of the Coulomb
interaction [Eq.\ (\ref{eq:matrix-element})] increase with increasing LL
index $L$, resulting in the observed spin dependence.

\begin{table}
\vspace{-.2cm}
\begin{center}
     \caption{   \label{tabl:sm2nem2kt} 
       Frank elastic constants $K_1$ and $K_3$,
%       [Eqs.\ (\ref{eq:F-nematic}), (\ref{eq:K1-K3})], 
       renormalized elastic constant $K$ 
%       [Eqs.\ (\ref{eq:F-xy}), (\ref{eq:KR})], 
       and KT disclination unbinding temperature 
%       [Eqs.\ (\ref{})-(\ref{})], 
       calculated for the experimental realization of 
       Ref.\ \protect\onlinecite{lilly99a}.
%        ($n_e = 2.67 \times 10^{11}$ cm$^{-2}$, 
%        $z_{\rm rms} = 58.3 {\; \rm \AA}$).  
	Note the characteristic
        oscillations with the spin index.
      }
   \vspace{0cm}
\begin{tabular}{|r|c||r|r|r||r|}
\hline
$\nu$ & $\sigma$
& $K_1\!$ (mK) & $K_3\!$ (mK) & $K\!$ (mK) & $T_{KT}\!$ (mK)  \\
\hline
9/2  & $\uparrow$ & 189 & 1030 & 610 &  206 \\
11/2 & $\downarrow$ &  144 & 783 & 463 &  156 \\ 
13/2 & $\uparrow$ &  192 & 1041 & 616 &  208 \\
15/2 & $\downarrow$ &  158 & 848 & 503 &  170 \\
17/2 & $\uparrow$ &  196 & 1034 & 615 &  208 \\
19/2 & $\downarrow$ &  167 & 875 & 521 &  176 \\
\hline
\end{tabular}
\end{center}
\vspace{-1.cm}
\end{table}

%-----------------------------------------------------------------
There are a couple of caveats which apply to our results. 
First,  we've left out the native anisotropy of the 
sample which tends to align the smectic structure (similar effects
arise from an in-plane component of the magnetic field). 
Uniaxial anisotropy will produce a term of the form
$B' (\partial_y u)^2$ in the smectic energy density; although 
the experiments indicate that $B'\ll B$, at sufficiently long
length scales (of order $\sqrt{K/B'}$) the anisotropy will 
dominate over the bending energy. In this case the dislocation 
energy diverges as the logarithm of the system size, and the transition
to the isotropic phase occurs through the unbinding of dislocations.
Second, as is customary in studies of smectics, 
we have dropped terms in the smectic free energy, 
Eq. (\ref{eq:LCmacro}), of ${\cal O}[q_x^2q_y^2]$.  To check the validity of
this truncation we have calculated the elastic coefficients $K'$ and $K''$, 
and find that while $K''>0$, it is possible for $K'$ to be 
negative \cite{abanov95}.  
This does not seem to cause any problems in the long wavelength limit, 
but it may change our estimates of the dislocation energy.  This issue
is currently under study \cite{wexler00b}.

In conclusion, we have mapped a 2DES with half-filled LLs to
a liquid crystal with smectic/nematic order at
short/long distances and which undergoes a KT
disclination unbinding transition, after which the system becomes
isotropic, as seen by transport measurements.  Without the use of any
fitting parameters we have obtained transition temperatures in
qualitative agreement with experimental evidence.  A particularly
robust feature is the spin dependence of the nematic elastic moduli and
transition temperature (Table \ref{tabl:sm2nem2kt}):  they are
larger for the lower spin sub-band ($\nu$ = 9/2, 13/2, 17/2).  While
precise experimental values for the transition temperatures have not
been established and the transition is rounded by disorder, the same
characteristic spin dependence is observed in the transport anisotropy 
$\rho_{xx}/\rho_{yy}$ \cite{lilly99a,eisenstein00b} (see also
Ref.\ \onlinecite{fradkin00}). 

We would like to acknowledge numerous helpful discussions with A.\
MacDonald, S.\ Girvin, E.\ Fradkin, J.\ Eisenstein, H.\ Fertig,
L.\ Radzihovsky, M.\ Lilly, M. Fogler and G.\ Vignale; ATD would also like to 
thank the Aspen Center for Physics for its hospitality during the completion
of this work.  This work was supported by the NSF Grant No.\
DMR-9978547 (ATD) and by the University of Missouri Research Board and
Research Council (CW).

%=================================================================
%=================================================================
%\newpage

%\end{thebibliography}

%=================================================================
%=================================================================
%=================================================================
%\end{multicols}
\end{document}